\documentclass[12pt]{iopart}
\usepackage{amssymb}
\usepackage{epsfig,psfrag}
\usepackage{graphicx}
\usepackage{pstricks}
\usepackage{color}
\usepackage{verbatim}

\begin{document}

\title{Comparative study of theoretical methods for nonequilibrium quantum
  transport}

\author{J Eckel$^1$, F Heidrich-Meisner$^2$, S G
  Jakobs$^{3,4}$, M Thorwart$^1$, M Pletyukhov$^{3,4}$ and R Egger$^5$}

\address{$^1$Freiburg Institute for Advanced Studies (FRIAS),
  Albert-Ludwigs-Universit\"at Freiburg, 79104 Freiburg, Germany}

\address{$^2$Department of Physics, Arnold Sommerfeld Center for Theoretical Physics,
and Center for NanoScience, Ludwig-Maximilians-Universit\"at M\"unchen, D-80333 M\"unchen, Germany}

\address{$^3$Institut f\"ur
  Theoretische Physik A, RWTH Aachen, 52056 Aachen, Germany}

\address{$^4$JARA -- Fundamentals of Future Information Technologies, RWTH Aachen, 52056 Aachen, Germany}

\address{$^5$Institut f\"ur Theoretische Physik,
  Heinrich-Heine-Universit\"at D\"usseldorf, 40225 D\"usseldorf,
  Germany} \date{\today}

\begin{abstract}
  We present a detailed comparison of three different
  methods designed to tackle nonequilibrium quantum
  transport, namely the functional renormalization group (fRG), the
  time-dependent density matrix renormalization group  (tDMRG), and the
  iterative summation of real-time path integrals (ISPI). For the
  nonequilibrium single-impurity Anderson model (including a Zeeman
  term at the impurity site), we demonstrate that the three methods
  are in quantitative agreement over a wide range of parameters at the
  particle-hole symmetric point as well as in the mixed-valence
  regime. We further compare these techniques with two quantum Monte Carlo
  approaches and the time-dependent numerical renormalization group method.  
\end{abstract}


\section{Introduction}
Nowadays possibilities of miniaturization down to the nanometer scale
allow for studying the single electron transport in ultra-small
devices such as, e.g., artificially designed molecular quantum dots
\cite{Weber02,Ruitenbeek97,Scheer04,Park,Park02} or nanotubes, with
nontrivial emergent physics such as Fermi-liquid behaviour in quantum
dots \cite{Potok07}.  Quantum many-body systems
driven to nonequilibrium tend to approach a stationary state being
distinct in character from their ground state properties
\cite{Abanin05,Mitra07}.  The details of the stationary state may in principle
depend on the way how the system is driven out of equilibrium and on
the character of the correlations in the system.  There is a broad
variety of interesting effects when a large bias voltage beyond the
regime of linear response is applied to the system due to interaction
effects or due to the nonequilibrium conditions
\cite{Koenig,Meir,Hershfield}.  These features range from coherent
transport properties, such as, e.g., resonant tunneling, to the Kondo
effect. A prominent nontrivial model to study the interplay of
correlation effects and quantum transport is the single impurity
Anderson model \cite{Anderson}. This model is believed to capture  essential features of
experiments with quantum dots \cite{gg,vanderwiel}. Applying a finite bias voltage across
the impurity region allows one to study nonequilibrium effects.  This regime 
is defined by the breakdown of the standard approach of linear response theory when 
the transport voltage becomes too large. 

There are different approximative approaches to deal with the
nonequilibrium situation. 
For example, quantum transport through a quantum dot in
the Kondo regime has been described theoretically by Fermi-liquid theory 
for the asymptotic low-energy regime \cite{Oguri01}, 
by means of interpolative perturbation schemes \cite{Aligia}, 
by exploiting the integrability of the model
\cite{Konik02,mehta06}, and in terms of the non-crossing approximation
\cite{meir93,wingreen94}.  Moreover, there exists a large class of
renormalization group approaches which are based on an expansion in the 
(renormalized) system-reservoir coupling and therefore allow for a treatment
of the weak-tunneling regime. For example, the Anderson dot in the charge 
fluctuation regime \cite{Schoeller00} and the time evolution of the spin-boson model \cite{Schoeller01}
have been initially considered in an early application 
of the real-time RG (RTRG) method in its real-time 
formulation \cite{Schoeller00a}. Nonequilibrium extensions
of the perturbative renormalization group \cite{Kaminski00,Rosch01,Rosch03} 
and of the flow equation approach \cite{Kehrein05} 
as well as the field-theoretical RG approach focusing on the Callan-Symanzik
equation \cite{doyon06}
have been applied 
to  the nonequilibrium Kondo effect. A more refined 
analytical perturbative scheme based on a reformulation of the 
RTRG in frequency space (RTRG-FS) \cite{Schoeller09} 
provides a systematic and self-consistent treatment of
observables \cite{Schoeller09a} and correlation functions \cite{Schoeller09b} 
in the nonequilibrium stationary state. 
Using RTRG-FS,
approximate analytic solutions have been proposed for the nonequilibrium Kondo model
for weak spin fluctuations \cite{Schoeller09a,Schoeller09b,Schoeller09c} and for the interacting resonant level model
for weak or strong charge fluctuations \cite{Andergassen09}.
Additionally, the RTRG approach also 
allows for a description of transient dynamics towards this state 
\cite{Schoeller01,Schoeller09c}. A complementary treatment of
transport properties through the Anderson dot can be based upon an 
expansion in the Coulomb interaction strength \cite{Fujii,rubio}. However,
bare perturbation theory appears to be insufficient for larger 
values of the interaction, and therefore, it should be combined with the concept of the  
renormalization group.  The latter prescribes how to renormalize interaction vertices in the
nonequilibrium case. This is achieved by
combining the Keldysh formalism with the functional renormalization
group (fRG) approach in the quantum field theoretical formulation
\cite{Gezzi07,Jakobs2,Jakobs09,Karrasch09}. Very recently, this
scheme has been improved by accounting for the frequency dependence of
the two-particle vertex \cite{Jakobs09}
in analogy to corresponding developments of
the equilibrium Matsubara fRG \cite{Hedden04,Karrasch08}.
As a result of this
improvement, the exponentially small scale of the Kondo temperature
has partly been observed, namely in the nonequilibrium Fermi-liquid coefficient, that is in the 
second derivative of the self-energy with respect to bias voltage. 
Moreover, fRG has been 
 applied to the study of transport
properties of a Kondo quantum dot as well \cite{Schmidt09}.

As an alternative to these approaches, numerical methods have
been advanced very recently to properly describe nonequilibrium
quantum transport. One important technique is the real-time quantum
Monte Carlo method (QMC) \cite{Egger00,Egger03}, which has been
extended to non-equilibrium
\cite{Muhlbacher08,Schmidt08,Werner09,Schiro09,Werner_arXiv}.
Although these results are valuable since they are numerically
exact, the related calculations are limited to short to intermediate
simulation times and the dynamical sign problem often does not allow one 
to reach the regime of the stationary current for low temperatures.
Recent real-time QMC simulations \cite{Werner_arXiv} seem to indicate that the
splitting of the Kondo resonance which was theoretically predicted
earlier in \cite{wingreen94,Fujii} is an artifact of the
perturbative approach, consistent with observations from tDMRG \cite{HeidrichMeisner09}.  
A novel
technique based on quantum Monte Carlo (QMC) simulations with complex
chemical potentials has been introduced in \cite{Han07}. This is
achieved by an analytic continuation to complex times by employing
Matsubara voltages. Then, standard equilibrium quantum Monte Carlo
techniques are used in this QMC variant \cite{Han07,Han_arXiv}.  Beyond raising
conceptually fundamental questions, a nonmonotoneous behaviour of the
conductance on the bias voltage has been reported \cite{Han07}, whose
origin remains unclear and might be attributed to the (arbitrary)
choice of the form of the scattering states on which the formalism
relies.

Moreover, the multi-layer multi-configuration time-dependent Hartree formalism 
has recently been introduced \cite{hartle08,wang09}, which is 
based on a clever decomposition of the overall
Fock space and which has been
applied to vibrationally coupled electron transport.
In addition, the numerical renormalization group (NRG) method has been
very successful in solving quantum impurity problems in equilibrium
for several years, see \cite{Bulla} for a topical review. This
approach has  been generalized to nonequilibrium by using a
discrete single-particle scattering basis
\cite{Anders05,Anders06,Anders07,Anders08,Roosen08,schmitt_arxiv}, 
resulting in the development of the time-dependent numerical
renormalization group (TD-NRG). In addition, in \cite{gross_arxiv}
and for certain limits of the single impurity Anderson model, a dynamical
steady state characterized by correlation induced current oscillations
is presented by means of the time dependent density functional theory (TD-DFT).  

Furthermore, the adaptive time-dependent density matrix
renormalization group method (tDMRG) that generally allows for the
simulation of the real-time evolution of pure states in strongly
correlated one-dimensional systems \cite{Schollwoeck04,White04} has
been applied to transport in nanostructures as well
\cite{White04,Schmitteckert04,schneider06,hassanieh06} and most
notably in the present context, to the calculation of current-voltage
characteristics of quantum dots
\cite{HeidrichMeisner09,Boulat_Saleur_Schmitteckert08,Dasilva08,kirino08}.
The steady-state currents are computed by taking suitable
time-averages over   time windows, in which initial transient
effects have disappeared already.

Finally, yet another numerical approach has been proposed that
is based on the deterministic iterative summation of real-time path
integrals (ISPI) \cite{Weiss08}. It determines a Keldysh generating
function for the time-dependent nonequilibrium current and yields
numerically exact results since the systematic errors within the
scheme are eliminated exactly. In a first step \cite{Weiss08}, ISPI
was developed for the single-impurity Anderson model and its validity
has been confirmed by a detailed comparison with various approximative
approaches within their range of applicability.

It is the aim of this work to compare  three of these methods, namely, tDMRG, fRG and
ISPI in detail, focusing on steady-state currents as a function of gate potential, voltage bias, 
temperature, and magnetic field. The goal is to establish their reliability by 
revealing an excellent quantitative agreement between results obtained with these techniques. 
We shall further check our results against those of other methods, namely the
TD-NRG \cite{Anders08} and two QMC schemes, first the one  from \cite{Han07,Han_arXiv} and second, the real-time 
QMC from \cite{Werner09,Werner_arXiv}.
In section \ref{sec:model}, we introduce 
 the single-impurity Anderson model, which is one the most fundamental models
for the description of quantum dots and which we will thus use to benchmark our methods. 
In section \ref{sec:methods}, we briefly review the
fRG, tDMRG and ISPI schemes. The comparison of the
respective numerical results and a discussion are contained in section \ref{sec:results}. 

\section{Model}
\label{sec:model}
In what follows we consider the single-impurity Anderson model \cite{Anderson} given by the
Hamiltonian ($\hbar=1$)
\begin{eqnarray}
\mathcal{H}&=&H_{dot}+H_{leads}+H_T \nonumber\\
&=&\sum_\sigma E_{0\sigma} \hat n_\sigma
+U \hat n_\uparrow \hat n_\downarrow + 
\sum_{kp\sigma}(\epsilon_{kp}-\mu_p)c^\dag_{kp\sigma}c_{kp\sigma}\nonumber\\
&&- \sum_{kp\sigma} \left[t_p c_{kp\sigma}^\dagger d_\sigma + h.c.\right].
\label{andham}
\end{eqnarray}
Here, $E_{0\sigma}=E_0+ \sigma B$ with $\sigma=\uparrow, \downarrow =
\pm$ is the energy of a single electron with spin $\sigma$ on the
dot. It can be varied by tuning a back gate voltage or by means of a
Zeeman term $\propto B$, under the assumption that the electron
dispersion in the leads is not influenced by the magnetic field. The
creation/annihilation operator for the dot electron is
$d_\sigma^\dagger/d_\sigma$, $\hat n_{\sigma}\equiv d^\dag_\sigma
d_\sigma$ has eigenvalues $n_\sigma=0, 1$, and the on-dot interaction
is $U$. The energies of the noninteracting electrons in the lead
$p=L/R=\pm$ with chemical potential $\mu_p=p eV/2$ are denoted by
$\epsilon_{kp}$ (with fermionic operators $c_{kp\sigma}$). The dot is
connected to the leads via the tunnel coupling $t_p$.

Within the fRG and the ISPI method, we assume the leads to be in thermal
equilibrium at temperature $T$ ($k_B=1$) and moreover take the standard wide-band limit with a constant
density of states, $\rho(\epsilon_F)$, around the Fermi energy, yielding the
hybridization $\Gamma_p=\pi\rho(\epsilon_F)|t_p|^2$. In this work, we
assume symmetric contacts, i.e., $\Gamma_L=\Gamma_R\equiv\Gamma/2$ and we
scale all quantities with respect to $\Gamma$. The
observable of interest is the symmetrized tunneling current $I=(I_L-I_R)/2$
with
\begin{equation}
\label{eq:current}
I(t)=-\frac{ie}{2}\sum_{kp\sigma}\left[pt_p \langle c_{kp\sigma}^\dag d_\sigma\rangle_t
-pt_p^* \langle d^\dag_\sigma c_{kp\sigma}\rangle_t
\right],
\end{equation}
where $I_p(t) = -e \dot{N}_p(t)$ with $N_p(t)= 
\langle \sum_{k\sigma}c_{kp\sigma}^\dagger 
c_{kp\sigma} \rangle_t$. The asymptotic long-time limit gives the stationary
steady-state dc current as $I=\lim_{t\to\infty}I(t)$.
\section{Methods}
\label{sec:methods}
Next, we briefly review the three techniques that we want to compare. 

\subsection{fRG}

A detailed explanation on how we apply the fRG to the single-impurity
Anderson model is given in reference~\cite{Jakobs2}. Here we only
sketch the main ideas.

Spectral as well as transport properties of the model can be comprehensively
described if an interaction contribution to the single-particle self-energy
$\Sigma$ is somehow established \cite{meir93}. A basic idea of the 
fRG approach
consists of a formulation of a flow equation for $\Sigma$ accounting for its 
full frequency dependence (therefore this RG approach is called {\it functional}). The starting
point of the flow is the noninteracting limit where the solution is known exactly. 
During the fRG flow the form of $\Sigma$ is continuously transformed,  
and the solution of an interacting problem is achieved when the flow terminates.

Technically, this procedure is set up by artificially
making the bare propagator $g$ depend on a flow parameter
$\lambda$. Being functionals of the bare propagator, the interacting
Green's and vertex functions acquire a dependence on $\lambda$ as well,
which is described by an infinite hierarchy of coupled flow equations
\cite{Salmhofer}. In this paper we focus on the flow of the
one-particle irreducible vertex functions \cite{Wetterich_Morris,
  Salmhofer2} which has proven to provide a successful approach to the
physics of several low-dimensional correlated electron problems
\cite{Metzner, Meden}.  The one-particle irreducible $n$-particle
vertex function can be defined diagrammatically as the sum of all
one-particle irreducible diagrams with $n$ amputated incoming lines
and $n$ amputated outgoing lines. Here we refer to 
diagrammatic perturbation theory on the Keldysh contour which results from an expansion in powers of the
two-particle interaction $U$.

We truncate the hierarchy of flow equations by neglecting the flow of
the three-particle vertex function. The flow equations for the
self-energy $\Sigma^\lambda$ and the two-particle vertex function
$\gamma^\lambda$ then read
\begin{eqnarray}
  \fl
  \label{eq:flow_self-energy}
  \frac{d}{d \lambda} \Sigma^{\lambda}_{1'|1} &=& -
  \frac{i}{2 \pi} \gamma^{\lambda}_{1'2'|12}\,
  S^{\lambda}_{2|2'}
  \\
  \fl
  \label{eq:flow_vertex}
  \frac{d}{d \lambda} \gamma^{\lambda}_{1'2'|12} &=&
  \frac{i}{2 \pi}
  \gamma^{\lambda}_{1'2'|34}\,S^{\lambda}_{3|3'}\,G^{\lambda}_{4|4'}\,
  \gamma^{\lambda}_{3'4'|12}
  + \frac{i}{2 \pi} \gamma^{\lambda}_{1'4'|32}
  \left[S^{\lambda}_{3|3'}\,G^{\lambda}_{4|4'} +
    G^{\lambda}_{3|3'}\,S^{\lambda}_{4|4'} \right]
  \gamma^{\lambda}_{3'2'|14}
  \nonumber
  \\
  \fl
  &&
  - \frac{i}{2 \pi} \gamma^{\lambda}_{1'3'|14}
  \left[S^{\lambda}_{3|3'}\,G^{\lambda}_{4|4'} +
    G^{\lambda}_{3|3'}\,S^{\lambda}_{4|4'} \right]
  \gamma^{\lambda}_{4'2'|32},
\end{eqnarray}
where indices occurring twice in a product imply summation over
state and Keldysh indices and integration over independent
frequencies. Furthermore, $G^\lambda$ is the full single-particle
propagator and $S^\lambda = - G^\lambda \left[d (g^\lambda)^{-1}/d
  \lambda\right] G^\lambda$ denotes the so-called single scale
propagator.

As flow parameter we use the hybridization constants $\Gamma_p$ by
setting
\begin{equation}
  \Gamma_p^\lambda = \Gamma_p + \lambda \frac{\Gamma_p}{\Gamma}, \quad
  p = L, R,
\end{equation}
where $\lambda$ flows from infinity to zero. This choice for the flow
parameter ensures the validity of causal properties and of the
fluctuation dissipation theorem in the special case of thermal
equilibrium \cite{Jakobs1}. The latter is important in order to
reproduce the Fermi liquid behaviour of the model near the
particle-hole symmetric point. We approximate the coupling of the
three channels in the flow equation~\eref{eq:flow_vertex} of the
two-particle vertex by reducing the influence of each channel onto the
other two to a renormalization of the interaction strength $U$. This
leads to a frequency dependence of the two-particle vertex of the form
\begin{equation}
\label{eq:two_part_vertex}
  \gamma^\lambda(\Pi,X,\Delta)
  \simeq 
  \overline v + \varphi^\lambda_1(\Pi) + \varphi^\lambda_2(X) +
  \varphi^\lambda_3(\Delta),
\end{equation}
where $\Pi=\omega_1+\omega_2=\omega_1'+\omega_2', X=\omega_2'-\omega_1=\omega_2-\omega_1', 
\Delta=\omega_1'-\omega_1=\omega_2-\omega_2'$ are bosonic frequencies corresponding to energy exchange in different channels. 
In particular, $\Pi$ refers to the particle-particle and $X$ to the exchange particle-hole channel, while 
$\Delta$ indicates the direct particle-hole channel. 
A similar approximation has been
investigated in studies of the single impurity Anderson model based
on the equilibrium Matsubara fRG \cite{Karrasch08}.
In \eref{eq:two_part_vertex},
$\overline v$ is the bare interaction vertex which is the initial value at the beginning of the flow, and 
$\varphi^\lambda_{1,2,3}$ are approximations to the corresponding parts of $\gamma$ produced 
by the flow equation, see \cite{Jakobs2}. 
This approximation considerably reduces the number of
sampling points in frequency space which are required for the
numerical solution of the flow equation.

\subsection{tDMRG}
The adaptive time-dependent DMRG method \cite{Schollwoeck04,White04} allows for the simulation of the time-evolution
of pure states $|\psi(t)\rangle$  at zero temperature (note that generalizations to mixed states and finite temperatures are possible, for a   review, see
\cite{schollwoeck05} and references therein). The basics of the technique itself are  described in the original
publications \cite{Schollwoeck04,White04} and the review  \cite{ulisteve}. In a nutshell, the key ingredient of the method is to
represent the wave-function in a truncated, but optimized basis instead of the exponentially large basis of the full Hilbert space (see \cite{schollwoeck05} for details).  
The application of tDMRG to transport in nanostructures was pioneered in \cite{HeidrichMeisner09,schneider06,hassanieh06,
Boulat_Saleur_Schmitteckert08,Dasilva08,kirino08}. We use the set-up of \cite{HeidrichMeisner09,hassanieh06}, which we shall briefly summarize.

In contrast to ISPI and fRG, we here apply tDMRG to a real-space representation of the
Hamiltonian, i.e., we replace the terms in \eref{andham} describing the leads and the
hybridization by
\begin{eqnarray}
H_{\mathrm{leads}}+H_{T} &=& - t_{p} \sum_{l=R,L;\sigma} (d_{\sigma}^{\dagger} c_{l,1,\sigma}+h.c.) \nonumber\\
        &&-\frac{W}{4}\sum_{n=1;l=R,L;\sigma}^{N_{L,R}} \,(c_{l,n\sigma}^{\dagger} c_{l,n+1 \sigma} ~+~h.c. )\,.\label{eq:ham}
\end{eqnarray}
Moreover, we work on finite systems with open boundary conditions. $N_{L(R)}$ is the number of sites
in the left (right) lead, with $N=N_L+N_R+1$.  The hopping matrix element in the leads is $W/4$, where $W$ is  the bandwidth.
In the present case of a semielliptical density of states, we obtain $\Gamma=2t_p^2/(W/4)$ for the hybridization.
Note that the leads can also be modeled with other density of states by a different choice of (typically site-dependent)
hopping matrix elements in the leads. A prominent example is the logarithmic discretization which is the standard choice 
in NRG, recently  used in some tDMRG simulations as well \cite{Dasilva08,guo08}. 

\begin{figure}[t!]
\begin{center}
\includegraphics[width=80mm]{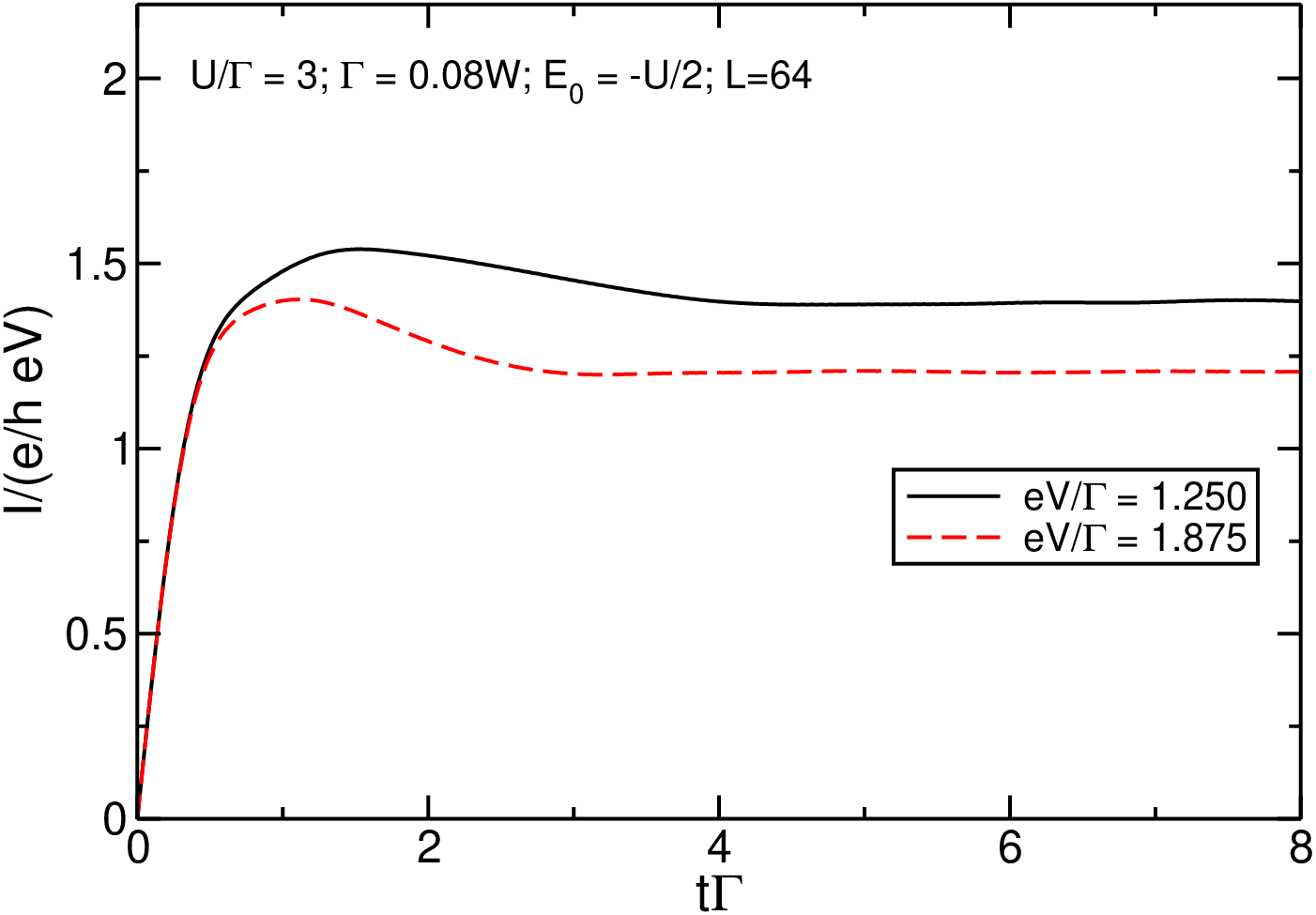}
\end{center}
\caption{\label{fig:0}Real-time evolution of the current $I(t)$ at the symmetric point
  $E_0=-U/2$ for $U/\Gamma=3$ for the tDMRG. Real-time results from the ISPI approach can be found in
\cite{Weiss08}, figure $5$ therein. 
} 
\end{figure}

A simulation then starts by computing the ground state of \eref{eq:ham}, defining the initial state $|\psi_0\rangle$, and then, at time
$t=0$, the system is driven out of equilibrium by applying on-site energies $\pm eV/2$ in the leads, mimicking
the bias
\begin{equation}
H_{\mathrm{bias}}=\frac{eV}{2}\sum_{n=1}^{N_R}
{n}_{R,n} -      \frac{eV}{2}\sum_{n=1}^{N_L}
{n}_{L,n}\,.
\end{equation}
For the time-evolution $|\psi(t)\rangle= \mbox{exp}(-iH t/\hbar)|\psi_0\rangle$, we use a  Trotter-Suzuki
breakup of the time-evolution operator with a time-step of $\delta t =0.1$  in units of  $1/(W/4)$.
During the time-evolution, we measure the symmetrized tunneling current  $ I(t)$, \eref{eq:current}.
Some typical results for the real-time data are shown in \fref{fig:0}.
As a function of time, $I(t)$ goes through a transient regime
before a steady-state is reached. 
The transient time, besides its dependence on $\Gamma$ 
(and, in the Kondo regime on the Kondo temperature $T_K$), is proportional to $1/V$ \cite{HeidrichMeisner09}.
The steady-state regime can be accessed in tDMRG simulations if
the typical transient times are smaller than $N/v_F$, where $v_F$ is the Fermi velocity in the leads.
At present, time scales of about $10/\Gamma$ can easily be reached at intermediate bias voltages as 
\fref{fig:0} illustrates (see also the discussion in \cite{HeidrichMeisner09}).
The final result of a simulation is, for our purpose, the steady-state current  that is computed by taking
time-averages over the real-time data. Methods to identify the quasi-steady-state and for finding the correct
time-windows for averaging are discussed in \cite{HeidrichMeisner09,schneider06} and we refer the reader to 
\cite{HeidrichMeisner09,schneider06,hershfield00} for a discussion of the time scales governing the transient regime.

The numerical errors inherent to the technique depend on two control parameters, first, the time-step $\delta t$ 
and second, the  discarded weight $\delta \rho$ \cite{schollwoeck05}. The latter is a measure of the truncation that the wave-function has been subjected to.
In the example studied here, the discarded weight is the relevant
quantity \cite{HeidrichMeisner09}.  Moreover, finite-size effects need to be taken into account and thus runs are carried out
for  $N \leq 96$ at small biases $V/t\lesssim 0.4$ and $N\leq 64$ at large biases.
All results presented in this work have been extrapolated in $1/N$. 

Unlike ground-state DMRG \cite{schollwoeck05}, there is no criterion that favourably limits the entanglement encoded in the time-evolved
wave-function (see also the discussion in \cite{HeidrichMeisner09} and references therein), rendering the simulation increasingly costly at long times and in particular, for large biases $V\gtrsim W/2$. Qualitatively, the larger the entanglement, the larger the dimension of the truncated basis set needs to be that is used to approximate $|\psi(t)\rangle$. This is at present the most crucial limitation of the method. Nevertheless,  
tDMRG has been very successful in yielding current-voltage characteristics for two of the most basic models
used to describe quantum dots, the interacting resonant level model  \cite{Boulat_Saleur_Schmitteckert08} and the single-impurity
Anderson model \cite{HeidrichMeisner09}. 
 
\subsection{ISPI}
The ISPI scheme \cite{Weiss08} is an in principle numerically exact method to deal with 
non-equilibrium transport through correlated quantum systems. The method is based on
the evaluation of a Keldysh generating function \cite{Keldysh}, which includes 
appropriate source terms to generate the observables of interest. Here, this is the 
nonequilibrium transport current. A real-time path integral 
\begin{equation}
Z[\eta]=\int \mathcal{D}\left[\prod_\sigma \bar{d}_\sigma , d_\sigma ,
\bar{c}_{kp\sigma} ,
c_{kp\sigma}\right]e^{iS[\bar{d}_\sigma, d_\sigma ,
\bar{c}_{kp\sigma}, c_{kp\sigma}]} ,
\end{equation} 
with Grassmann fields $(\bar{d},d,\bar{c},c)$ (here we use the same
symbols for the Grassmann fields and the fermionic operators in
(\ref{andham})) is constructed for the generating function. The action $S$
corresponds to the Hamiltonian (\ref{andham}) and an external source
term is added to the action that allows for computing the dc current
at measurement time $t_m$ via
$I(t_m)=\left.-i\frac{\partial}{\partial\eta}\ln
  Z[\eta]\right|_{\eta=0}$. The full real-time evolution is computed
by successively inserting $t_m$ at different Trotter slices, e.g. in \cite{Weiss08}, figure 5 therein.
Next, the path integral is Trotter
discretized along the Keldysh contour.  The interaction ($U$-) term is
decoupled via a discrete Hubbard-Stratonovich transformation on each single Trotter 
slice, by which an auxiliary ``spin''-field is introduced. For the single-impurity
Anderson model, it can assume the values $\pm 1$. Then, all fermionic
fields can be integrated out analytically. The remaining task is to 
perform the summation over all configurations of the (bosonic) 
auxiliary fields, reminiscent of a discrete path
summation. Hence, in
\begin{equation}
  Z[\eta]=\sum_{\{s\}}\prod_\sigma (-i\det G_\sigma^{-1}[\{ s \},\eta])  ,
\label{pathsum}
\end{equation}
(with the total effective Keldysh Green's function $G_\sigma^{-1}$), the summation 
over  all configurations ${s}$ 
is performed numerically in a deterministic way. The key ingredient for this is the 
fact that the fermionic environment (the leads) induces  non-local 
correlations in time which decay exponentially in the long-time limit for any finite
temperature. This implies the existence of a characteristic memory time 
up to which all correlations
are fully taken into account and beyond which they can be safely neglected 
for larger times,
similar to the case of the spin-boson model \cite{eckel06}. This allows us
to disentangle the complete path sum and to 
construct an iterative scheme to evaluate the generating function. By means of a properly
chosen extrapolation procedure, both systematic numerical 
errors of the scheme, namely the Trotter error and the 
error due to the finite memory-time, are eliminated. If the extrapolation scheme will be convergent, one is left with the desired
numerically exact values for the observable at a fixed measurement time. 
Convergence is generally problematic when both $T$ and $V$ approach zero.
For more details, we refer the reader to \cite{Weiss08}.  


\section{Results and Discussion}
\label{sec:results}
For the comparison of the three approaches, we compute the symmetrized
steady-state current \eref{eq:current}. A
comparison of the fRG and the tDMRG results has already been shown in
\cite{HeidrichMeisner09} for zero temperature and both at particle-hole symmetry
and in the mixed valence regime. Consistency between these two methods
could be demonstrated for $U/\Gamma \leq 8$ and $eV> T_K$, where $T_K$
is the Kondo temperature. We now set out to include ISPI results into this comparison, starting 
with the zero temperature transport at the symmetric point $E_0=-U/2$ at finite on-dot interaction $U$.
We shall then  extend our study to the mixed valence regime $E_0\neq-U/2$, and end up with including a finite
magnetic field 
and finite temperatures, presenting new data from all three methods.

\begin{figure}[t!]
\begin{center}
\includegraphics[width=80mm]{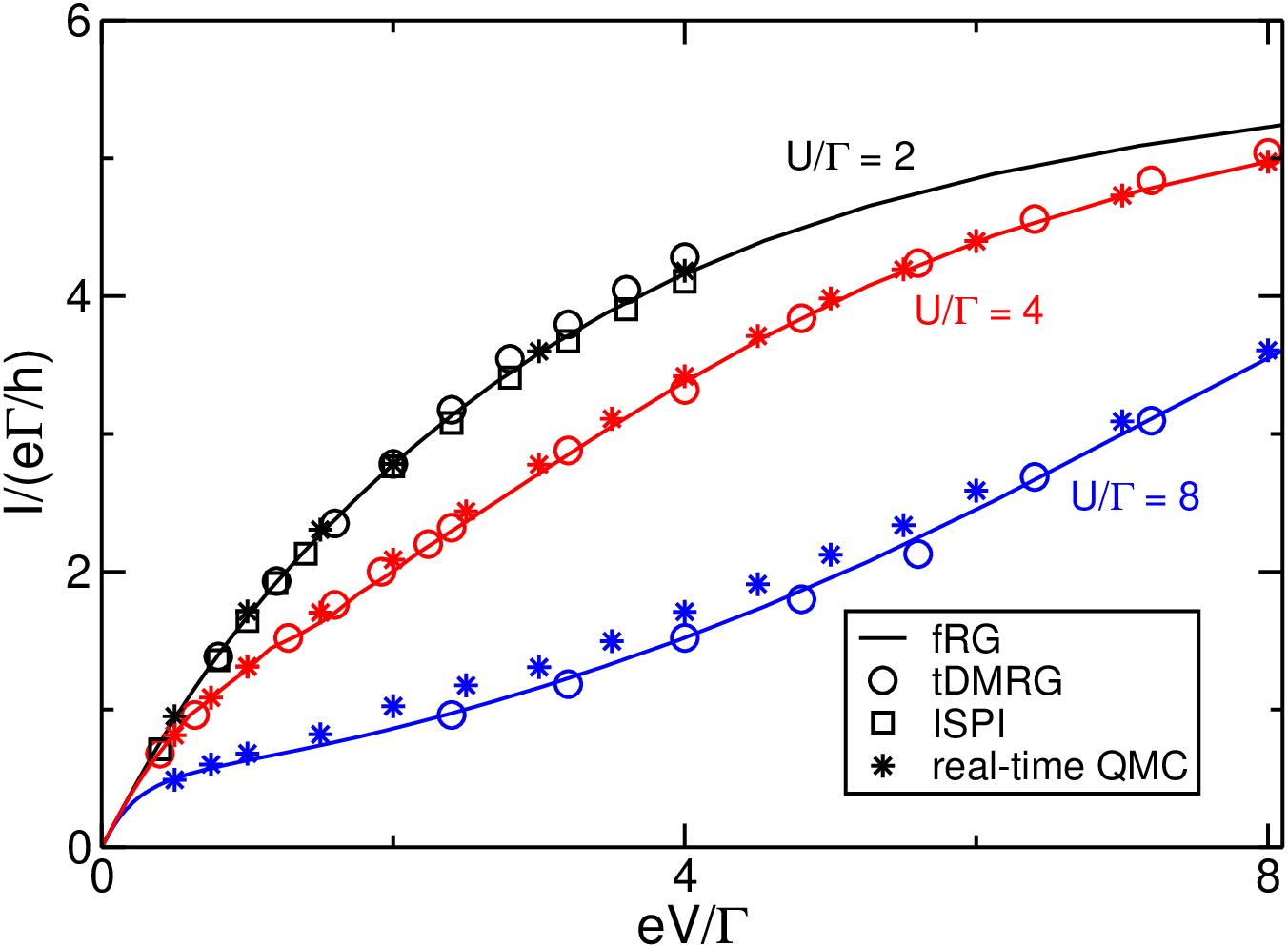}
\end{center}
\caption{\label{fig:1}Steady-state current $I$ at the symmetric point
  $E_0=-U/2$ for $U/\Gamma=2$ as a function of the applied bias voltage
  $eV/\Gamma$.  The tDMRG results are taken from \cite{HeidrichMeisner09} and 
were computed for $U=W/8$ ($\Gamma$ chosen correspondingly) and extrapolated in the inverse system size.
The real-time QMC data are taken from \cite{Werner09} ($U/\Gamma=2$) and \cite{Werner_arXiv}
($U/\Gamma=4$ and $8$). 
In the ISPI approach, we set the  temperature to $T=0.1\Gamma$. 
} 
\end{figure}

In \fref{fig:1}, we show the
results for the particle-hole symmetric point, $E_0=-U/2$, of the
single-impurity Anderson model. For the case of $U/\Gamma=2$ we find
an excellent agreement between the three methods and with the results from
the nonequilibrium real-time QMC approach of \cite{Werner09}. The fRG is
expected to be very reliable for $U/\Gamma=2$ and we can thus use fRG to   benchmark the other methods in this limit. 
Generally (the same applies to the
following figures as well), the
deviations at large bias voltages between tDMRG on the one hand and fRG and ISPI on the other hand are due to the finite
bandwidth used in the tDMRG calculations, as compared to the wide-band limit taken in the 
other two techniques (see \cite{HeidrichMeisner09} for details). For the large bias voltage (around
$eV/\Gamma\sim 4$) tDMRG and
ISPI deviate from each other by up to $4\%$.   
The high accuracy of the
ISPI data is illustrated by our findings shown in \fref{fig:1},
where the method agrees very well with both the fRG and the tDMRG, to be specific deviations are below $1.5\%$.
For $U/\Gamma=4$ we find good agreement of the fRG and the tDMRG with the real-time
QMC approach from \cite{Werner_arXiv}. In the case of $U/\Gamma=8$, fRG and tDMRG yield very similar results
whereas
the real-time QMC gives a slightly larger current. Yet these deviations are within the numerical error inherent to the techniques, see e.g. \cite{HeidrichMeisner09}.

\begin{figure}[t!]
\begin{center}
\includegraphics[width=80mm]{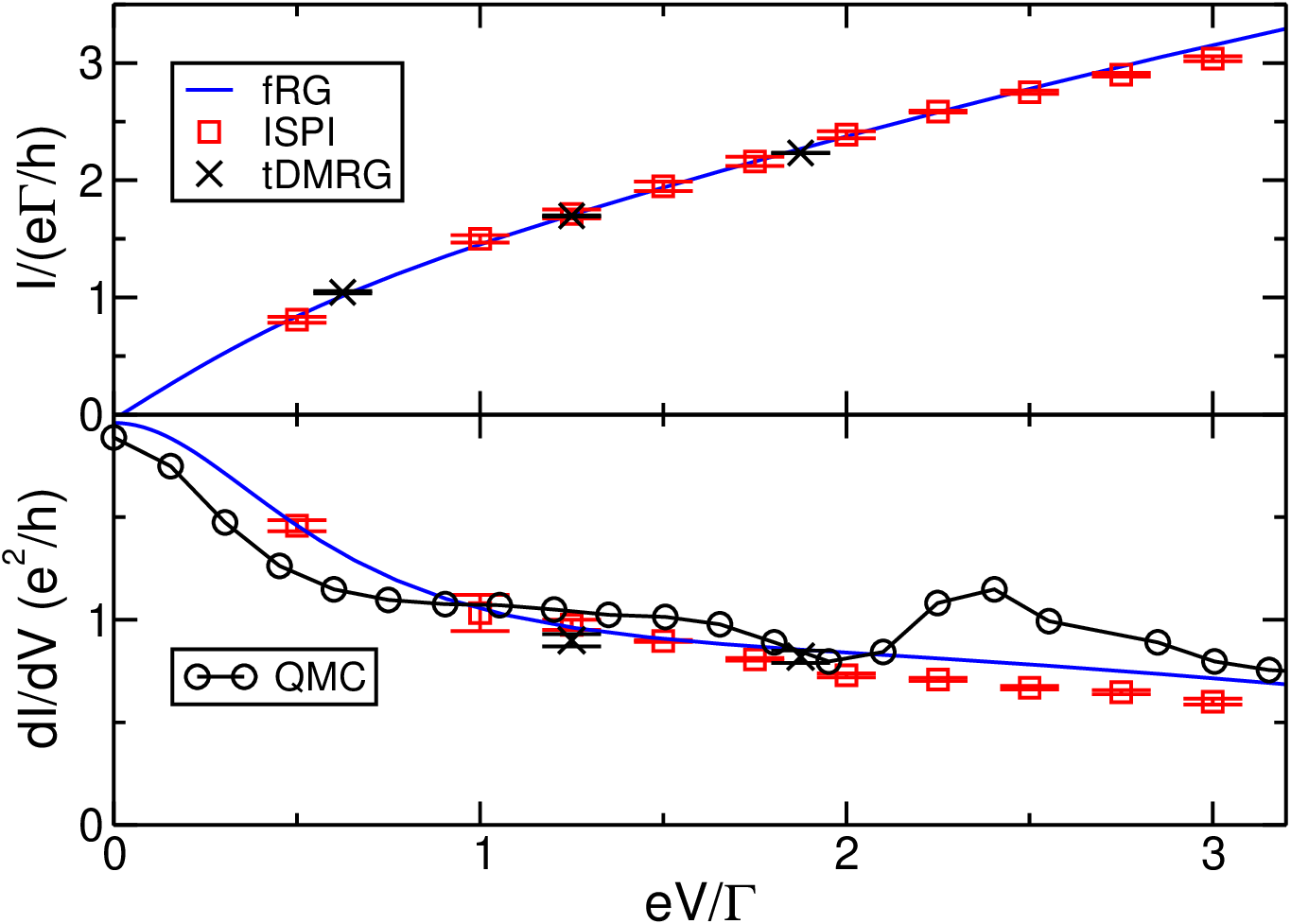}
\end{center}
\caption{\label{fig:5} Current-voltage characteristics (top) and  
differential conductance (bottom) for $U/\Gamma=3$. The 
temperature is $T=\Gamma/25.6$ and the remaining parameters are the same
as in \fref{fig:1}. The circles represent the results from a QMC method using
an analytic continuation to imaginary Matsubara voltages; 
\cite {Han07} (figure 2 therein).} 
\end{figure}

Next, we compare the results of fRG, ISPI and tDMRG at intermediate 
interactions and set $U/\Gamma=3$ 
(ISPI and fRG are computed at $T=\Gamma/25.6$). The  respective
current-voltage characteristics $I(V)$ 
is shown in \fref{fig:5} (top). We find a nice agreement over a large range of 
$V$, while again, small deviations between ISPI and fRG occur at very 
large voltages. In \fref{fig:5} (bottom), we compare the differential 
conductance obtained from our three methods with the result of Han and Heary 
 \cite{Han07} (see figure 2 in that work). 
We numerically approximate $dI/dV \approx [I(V+\Delta V)] -I(V)/\Delta V$
by choosing a sufficiently small $\Delta V$.

Strong differences between the QMC result and our three techniques occur and the non-monotonous 
features of the QMC data are not reproduced in our simulations. Moreover, the results of fRG and ISPI coincide 
at small to intermediate voltages, while 
some deviations occur at larger voltage \footnote{More recent versions of the method of \cite{Han07} have
given somewhat better agreement \cite{Han_privat}.
An extensive  discussion of this method has recently been presented in \cite{dirks_arxiv}.}. 
Within the error bars of the numerical differentiation and also taking into account uncertainties
in the bare $I(V)$,   tDMRG yields results for $dI/dV$ that are consistent with those from fRG and ISPI. 

\begin{figure}[t!]
\begin{center}
\includegraphics[width=80mm]{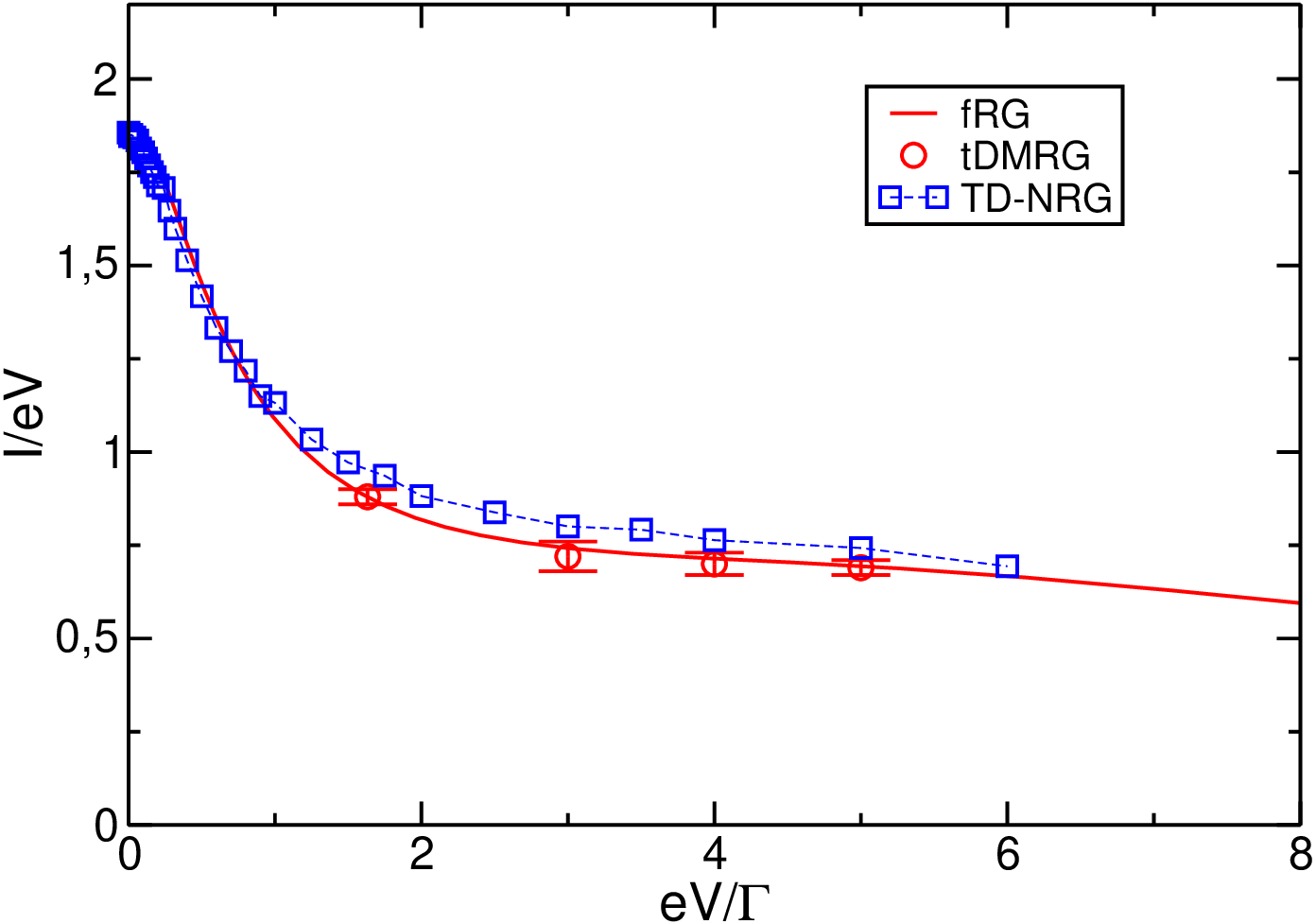}
\end{center}
\caption{\label{fig:6}
Comparison of the steady-state current $I$ divided by $eV$ for $U/\Gamma=5$
for the fRG, tDMRG and the TD-NRG \cite{Anders08}. The remaining parameters are the same
as in \fref{fig:1}. } 
\end{figure}

In a next step, we further increase the Coulomb interaction and 
choose $U/\Gamma=5$, for which TD-NRG results are available \cite{Anders08}. Using ISPI in its present formulation, this choice for $U/\Gamma$
does not yet allow us to obtain  converged results. In \fref{fig:6}, 
we show the results for $I$ from fRG and  tDMRG, together with 
 the TD-NRG data by Anders \cite{Anders08}. We find an excellent agreement between fRG and tDMRG 
for $eV \geq 1.75 \Gamma$. The TD-NRG data are very similar to fRG
at  small bias voltages, while TD-NRG gives larger values for the steady-state current 
 at large values of $V$ than the other two methods. The origin of this discrepancy is, at present, not understood,
yet we have reason to speculate that discretization errors may play a role here. tDMRG simulations
for an Anderson impurity coupled to Wilson leads, i.e., applied  to the same Hamiltonian that NRG treats,
also result in too large currents at $eV > T_K $ \cite{Dasilva08}. Using improved discretization schemes
might provide a remedy for this problem and should be tested in future work (see, e.g.,  \cite{zitko09,zitko09b,weichselbaum09}).

\begin{figure}[t!]
\begin{center}
\includegraphics[width=80mm]{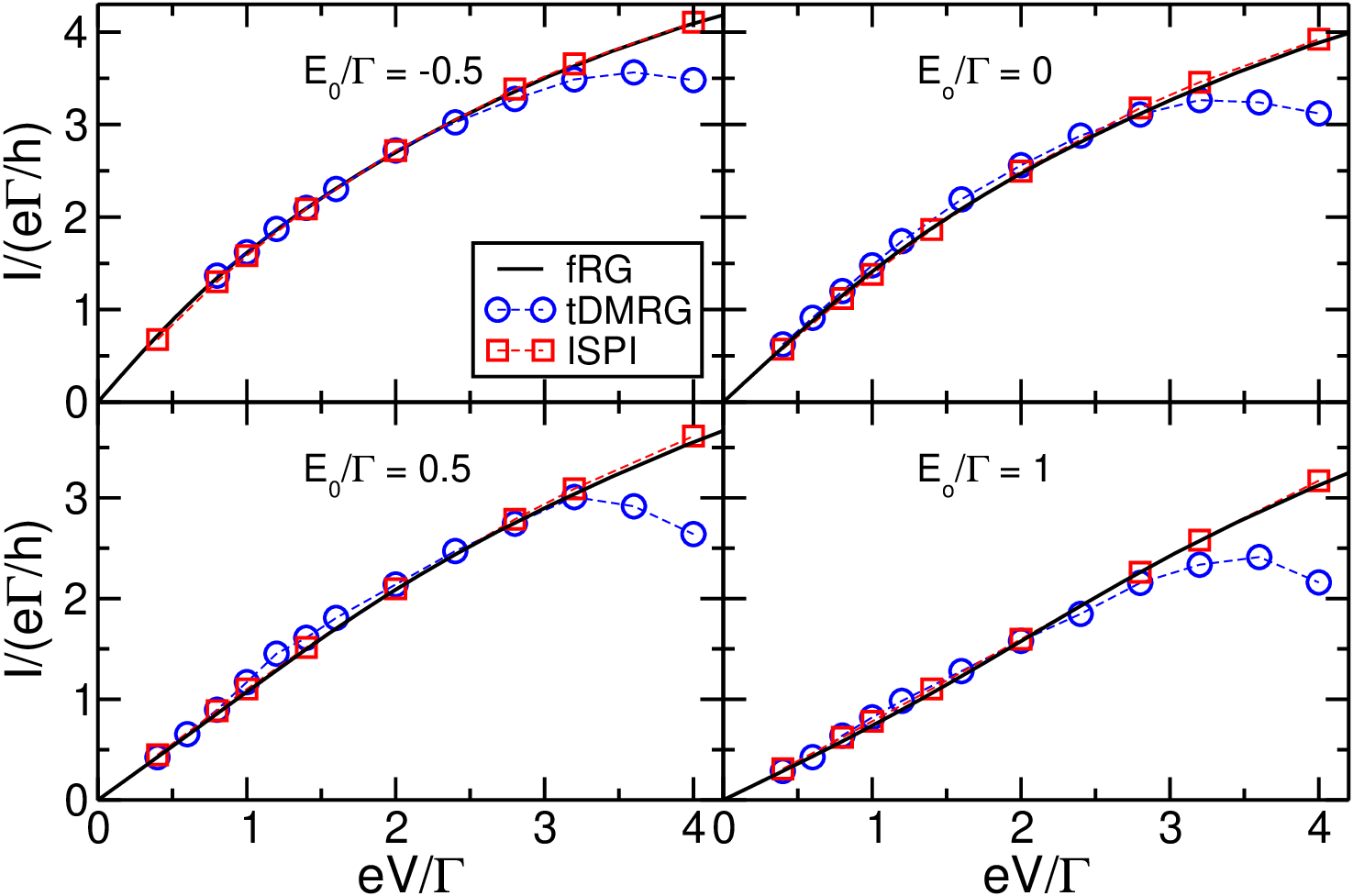}
\end{center}
\caption{\label{fig:2}Steady-state current for the mixed valence regime for
  different single electron energies $E_0$ with $U/\Gamma=2$. 
   The remaining parameters are the same as in \fref{fig:1}. The tDMRG data are taken from \cite{HeidrichMeisner09}.  
} 
\end{figure}

Next, we address the mixed valence regime for the case of
$U/\Gamma=2$. In \fref{fig:2}, we show the steady-state current $I$ for
different single electron energies $E_0\not= -U/2$. The results from  fRG and ISPI
match perfectly from small bias voltages $eV/\Gamma\sim 0.2$ 
up to the strong non-equilibrium regime. 
Note that by construction, it becomes increasingly cumbersome (and finally impossible) to 
obtain converged ISPI results in the limit of vanishing bias voltages and low temperatures 
\cite{Weiss08}, since then the correlations do not decay sufficiently fast enough  
to be truncated. 
In the present set-up, tDMRG has a tendency to overestimate the 
currents in the mixed valence regime, for more details, see \cite{HeidrichMeisner09}. This, we believe is the reason for slight deviations from
the other two methods  (see $eV \approx 1.5\Gamma$, see \fref{fig:2}, lower left panel). 
Overall, the agreement between the three methods   
is obviously still very good, even  away from the symmetric point.

\begin{figure}[t!]
\begin{center}
\includegraphics[width=80mm]{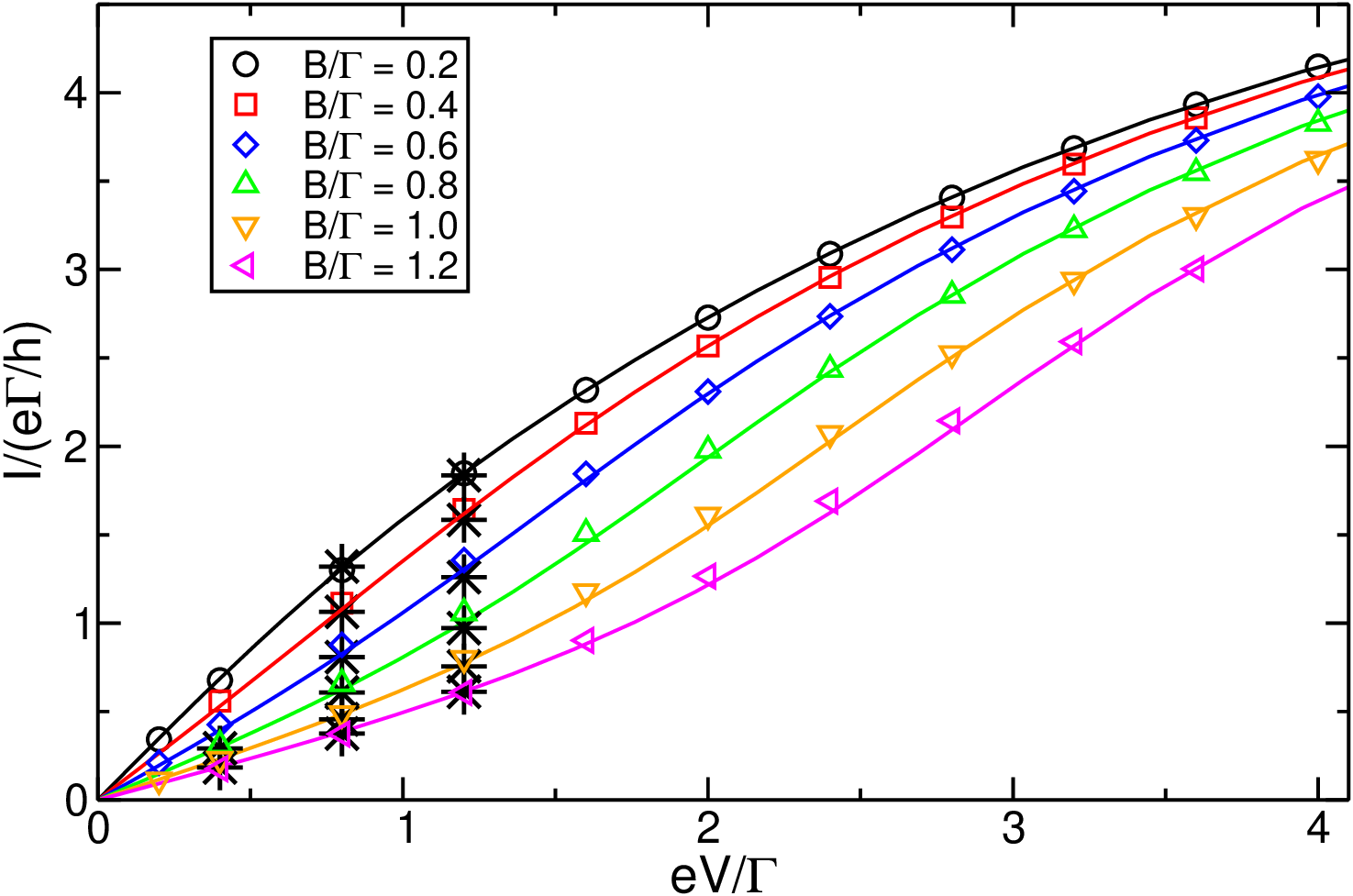}
\end{center}
\caption{\label{fig:3}
Steady-state current $I$ for finite magnetic field $B$, remaining parameters are the same as in
\fref{fig:1}. The solid lines are
the fRG results corresponding to the symbols of the ISPI results and the corresponding
tDMRG results are marked with the black stars. 
}    
\end{figure}

Moreover, we consider the case of a finite magnetic field $B$ at
the symmetric point, see \fref{fig:3}. We find an excellent agreement between fRG
and ISPI over the full range of bias voltage (deviations are below $1\%$) and  the tDMRG data
agree well with the two other methods for $eV \leq 1.5\Gamma$,
where we can mimic the wide band limit by keeping $\Gamma,U,B,eV \ll W $.
 
\begin{figure}[t!]
\begin{center}
\includegraphics[width=80mm]{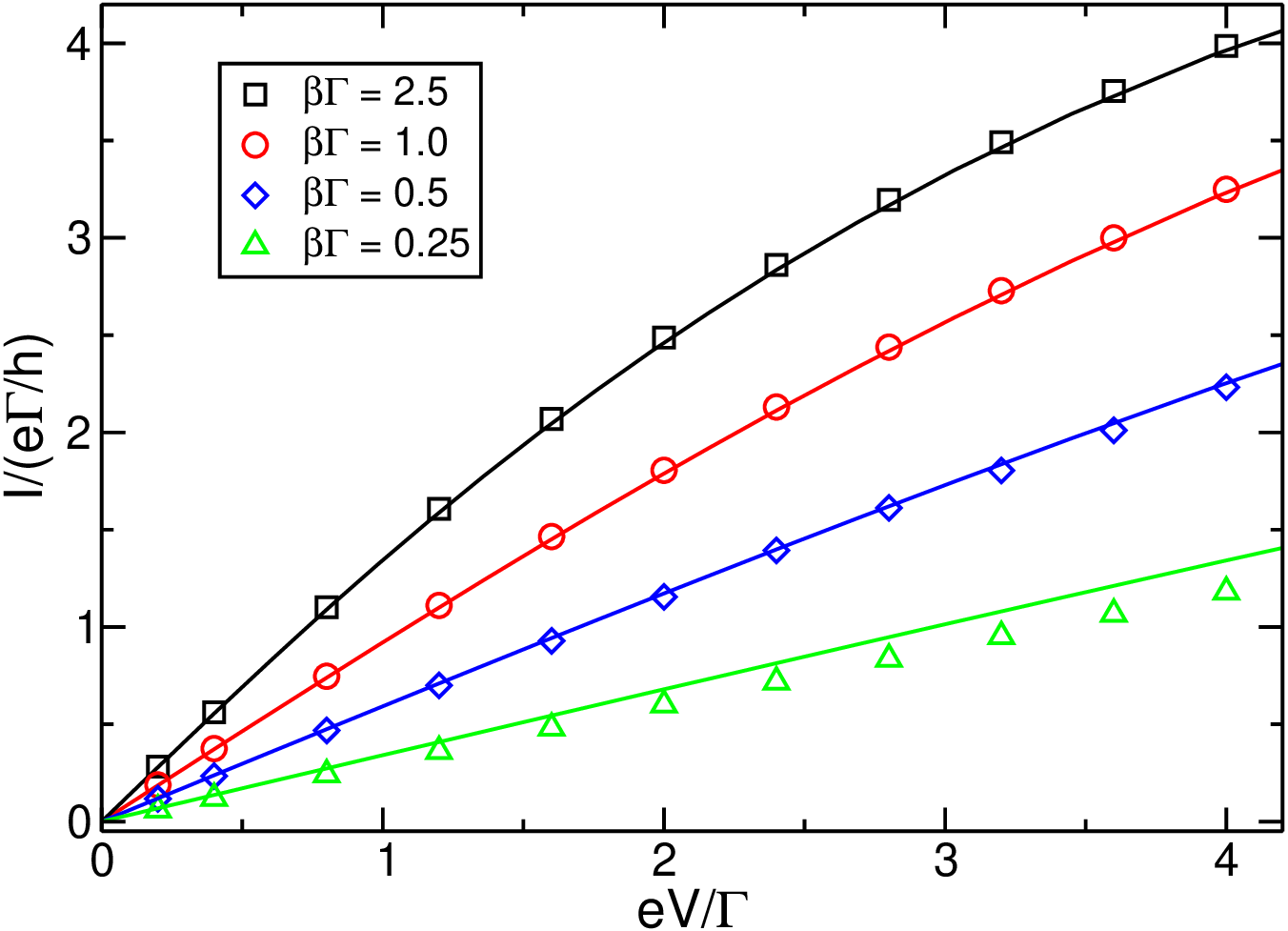}
\end{center}
\caption{\label{fig:4}
Temperature dependence of the steady-state current. The solid lines correspond to the
fRG results, the symbols denote the ISPI results. The remaining parameters are the same as in \fref{fig:1},
with $\beta=1/T$ being the inverse temperature.} 
\end{figure}

Finally, in the case of  finite temperatures $T$ (see \fref{fig:4}), for which at present no tDMRG is available yet,  
we again find
a remarkable agreement between fRG and ISPI from the equilibrium up to the strong
non-equilibrium case, with deviations below $1.5\%$. Only at large temperatures both methods tend to disagree
slightly for larger bias voltages and deviate up to $10\%$.  We note that for increasing
bias voltages, the fRG approach becomes more and more reliable and even 
asymptotically exact as $V \to \infty$. ISPI also profits from growing 
bias voltages as they cut off the time non-local correlations 
in the leads' Green's functions more and more efficiently.

In the remaining part, we wish to point out advantages and disadvantages of the
three techniques this work was mainly concerned with, the fRG,
tDMRG, and ISPI. fRG is a computationally rather cheap tool to compute
current-voltage characteristics over the entire bias range and at
practically all temperatures. The approximations taken in this method
have been estimated to be valid up to $U/\Gamma\approx 6$.
In particular, in \cite{Jakobs09} it has been shown that 
they yield a quantitatively accurate description of the linear conductance as a function 
of gate voltage, temperature and magnetic field; 
good descriptions of equilibrium properties of
the model have already been observed in Matsubara implementations
of the fRG \cite{Karrasch08,Karrasch06}.
  Concerning nonequilibrium properties, 
we obtain very good agreement for the nonlinear current and differential 
conductance for $U/\Gamma \leq 8$ with tDMRG data 
close to the most critical point $E_0 - U/2=T= B = 0$ \cite{HeidrichMeisner09}. 
For sufficiently low (but still finite) temperatures,  ISPI yields results that 
also compare well with fRG.
Within the Keldysh fRG framework, we are able to recover \cite{Jakobs09} 
the nonequilibrium Fermi-liquid relation \cite{Oguri01} and identify an
exponentially small energy scale of the Kondo temperature in the corresponding
Fermi-liquid coefficient which is the second derivative of the self-energy 
with respect to bias voltage (although, true Kondo physics for large $U/\Gamma$ cannot 
be described in general \cite{Jakobs09}). However, the reliability of fRG results at 
larger values of $U/\Gamma \gtrsim 6$ obtained in the scheme truncated at 
the level of the three-particle vertex is unknown at present, since 
the justification to truncate the fRG hierarchy of equations can only 
be given by perturbative arguments. In the single-impurity Anderson model, it is not a priori clear 
that the contribution of a three-particle vertex is negligible at large $U$. This 
constitutes an interesting subject for future research.  

Using tDMRG, one can access the finite voltage range with $eV\gtrsim T_K$, 
as has been demonstrated for $U/\Gamma \lesssim 8$ in \cite{HeidrichMeisner09} and in this work,
while the full current-voltage characteristics can be obtained  
at  small $U/\Gamma \sim 4$. The main limitation of the method is the entanglement growth, rendering  
 long time-scales difficult to access at large voltages,
which at present excludes an analysis of the deep Kondo regime.
A partial solution to this problem is the use of leads with a logarithmic discretization,
and then, the correct result is obtained for the linear conductance and $U/\Gamma \lesssim 7$
\cite{Dasilva08}. 
tDMRG results can be rendered quite accurate as well,
yet since the method requires an exponentially increasing computational time
both for reaching long time scales and to obtain very accurate data, in practice, one has to 
accept a finite numerical error, which can be estimated 
by tuning control parameters. The greatest advantage of tDMRG is that 
existing codes can be directly applied to both complex 
interacting structures as well as interacting leads \cite{feiguin08}.

Regarding the ISPI scheme, an important advantage is that 
whenever it converges with respect to its internal control parameters, 
numerically exact results are obtained, where no sign problem 
restricts the accessible simulation times and the steady-state regime
can directly be reached.
In the present implementation, the ISPI scheme converges for sufficiently
high temperature \emph{or} sufficiently large bias voltage. 
The non-local time correlations induced by integrating out the 
fermions in the leads within this approach become long-ranged only 
when $T$ and $V$ both approach zero. Such a slow decay of 
time correlations necessitates a long memory time, i.e.,
one has to take into account more and more memory slices within the
path sum in \eref{pathsum}, where the computational effort
scales exponentially in the number of memory slices. 
This problem becomes more severe for
increasing $U$, and an exploration of the deep Kondo limit
remains challenging at present, since the long-range correlations in time are
essential for taking into account characteristic features of the Kondo limit. 
Further improvements of the scheme are under study
to allow more efficient calculations in the large-$U$ limit.
However, as we have shown above, the present ISPI implementation
is able to provide highly accurate and reliable results for
$U/\Gamma \lesssim 3$, and the zero temperature limit 
can be reached for voltages $V\gtrsim 0.1 \Gamma$. 
Moreover ISPI allows for accessing the full real-time characteristics
of observables of interest up to in principle asymptotically long times in
contrast to real-time QMC approaches \cite{Muhlbacher08} where the error bars
increase due to the sign problem for long times.
It should
also be stressed that ISPI is particularly well suited to 
treat the limit of intermediate-to-high voltages, where 
alternative methods often run into difficulties.

To summarize, we presented a detailed comparison of theoretical methods 
designed to model nonequilibrium transport, taking the example 
of the single-impurity Anderson model. A very encouraging quantitative agreement 
between the fRG, ISPI, and tDMRG could be established, suggesting that 
a well-equipped toolbox is available to study nonequilibrium transport
in nanostructures. We can also conclude that our methods fair well
against alternative developments, namely TD-NRG \cite{Anders08} and two QMC variants \cite{Werner09,Werner_arXiv,Han07},
of course keeping in mind that not all schemes can be applied in all parameter regimes.
In conclusion, each separate approach to nonequilibrium 
quantum transport bears its own right since different 
aspects of the real-time dynamics of correlated 
nonequilibrium quantum systems can be learned from 
these different approaches. 

\ack SGJ and MP
thank H. Schoeller for numerous valuable discussions on the
application of the nonequilibrium fRG,
and C. Karrasch and V. Meden for
sharing their experience in treating the single impurity Anderson model with fRG
methods.
JE, MT and RE thank S. Raub for useful technical help with the
optimization of the numerical code of the ISPI scheme, and S. Weiss, D. Becker and 
R. H\"utzen for useful discussions. 
FHM acknowledges fruitful discussions with A. Feiguin and E. Dagotto and
is indebted to E. Dagotto for granting us computing time at his group's facilities at the University of Tennessee
at Knoxville, where the tDMRG simulations were carried out.
We further thank F. Anders and the authors of \cite{Werner09} and \cite{Werner_arXiv} for sending us their results
and V. Meden for comments on a previous version of the text.
This work was supported by
the DFG Priority Program 1243, by the Excellence Initiative of the
German Federal and State Governments and by the DFG Forschergruppe 723.  
Computational time from the ZIM at the
Heinrich-Heine-Universit\"at D\"usseldorf is also acknowledged.

\newpage



\end{document}